\documentclass[twocolumn,showpacs,preprintnumbers,amsmath,amssymb,superscriptaddress,10pt,aps,prb]{revtex4-1}
\usepackage{epsfig,amsopn}
\usepackage{graphicx}
\usepackage{epstopdf}
\usepackage{amsmath,amssymb}
\usepackage{amsthm}
\usepackage{enumerate}
\usepackage{xcolor}
\newcommand{\angstrom}{\textup{\AA}}

\begin{document}
\title{Dynamical conductivity of the Fermi arc and the Volkov-Pankratov states on the surface of Weyl semimetals}

\author{Dibya Kanti Mukherjee}
\affiliation{Laboratoire de Physique des Solides, CNRS UMR 8502,
Universit\'e Paris-Sud, 91405 Orsay Cedex, France}
\author{David Carpentier}
\affiliation{Universit\'e de Lyon, ENS de Lyon, Universit\'e Claude Bernard, CNRS, Laboratoire de Physique, F-69342 Lyon, France}
\author{Mark Oliver Goerbig}
\affiliation{Laboratoire de Physique des Solides, CNRS UMR 8502, Universit\'e Paris-Sud, 91405 Orsay Cedex, France}
  
\begin{abstract}
  Weyl semimetals are known to host massless surface states called Fermi arcs. These Fermi arcs are the manifestation of the bulk-boundary correspondence in topological matter and thus are analogous to the topological chiral surface states of topological insulators. It has been shown that the latter, depending on the smoothness of the surface, host massive Volkov-Pankratov states that coexist with the chiral ones. Here, we investigate these VP states in the framework of Weyl semimetals, namely their density of states and magneto-optical response. We find the selection rules corresponding to optical transitions which lead to anisotropic responses to external fields. In the presence of a magnetic field parallel to the interface, the selection rules and hence the poles of the response functions are mixed.
\end{abstract}
	
\maketitle
	
\section{Introduction}

One of the most striking features of topological materials is the bulk-boundary correspondence. These materials typically have a band gap and nontrivial invariants associated with the bulk. The invariants do not depend on the precise values of the system parameters and are robust to smooth deformations of the Hamiltonian as long as the band gap is not closed\cite{Bernevig2013,Hasan2010,Qi2011}. In fact, the only way to change the topological invariants is to close the band gap, thus creating a massless state. This is what happens at a topological heterojunction\cite{Tchoumakov2017a}, where one side has a topological system and the other side of the junction has a trivial system, which may even represent the vacuum. Such heterojunctions thus host one or more massless edge or surface states which are toplogically protected by the bulk boundary correspondence.

Already in the 1980-s, before the advent of topological matter, it was shown by Volkov and Pankratov\cite{Pankratov1987} that along with the massless surface states, massive states can also exist in such heterojunctions if the junction is sufficiently smooth. Indeed, the number of such states was estimated to be $\sim \ell/\xi$, here $\ell$ is the length over which the band gap vanishes and thus characterizes the smoothness of the junction. The intrinsic length scale of the states, $\xi=\hbar\Delta/v$ is given by the ratio between the band gap and the characteristic velocity of the system. Transport signatures of these states under high electric field have been observed experimentally on the surface of HgTe systems\cite{Inhofer2017}.

\begin{center}
  \begin{figure}
    \includegraphics[width=0.45\textwidth]{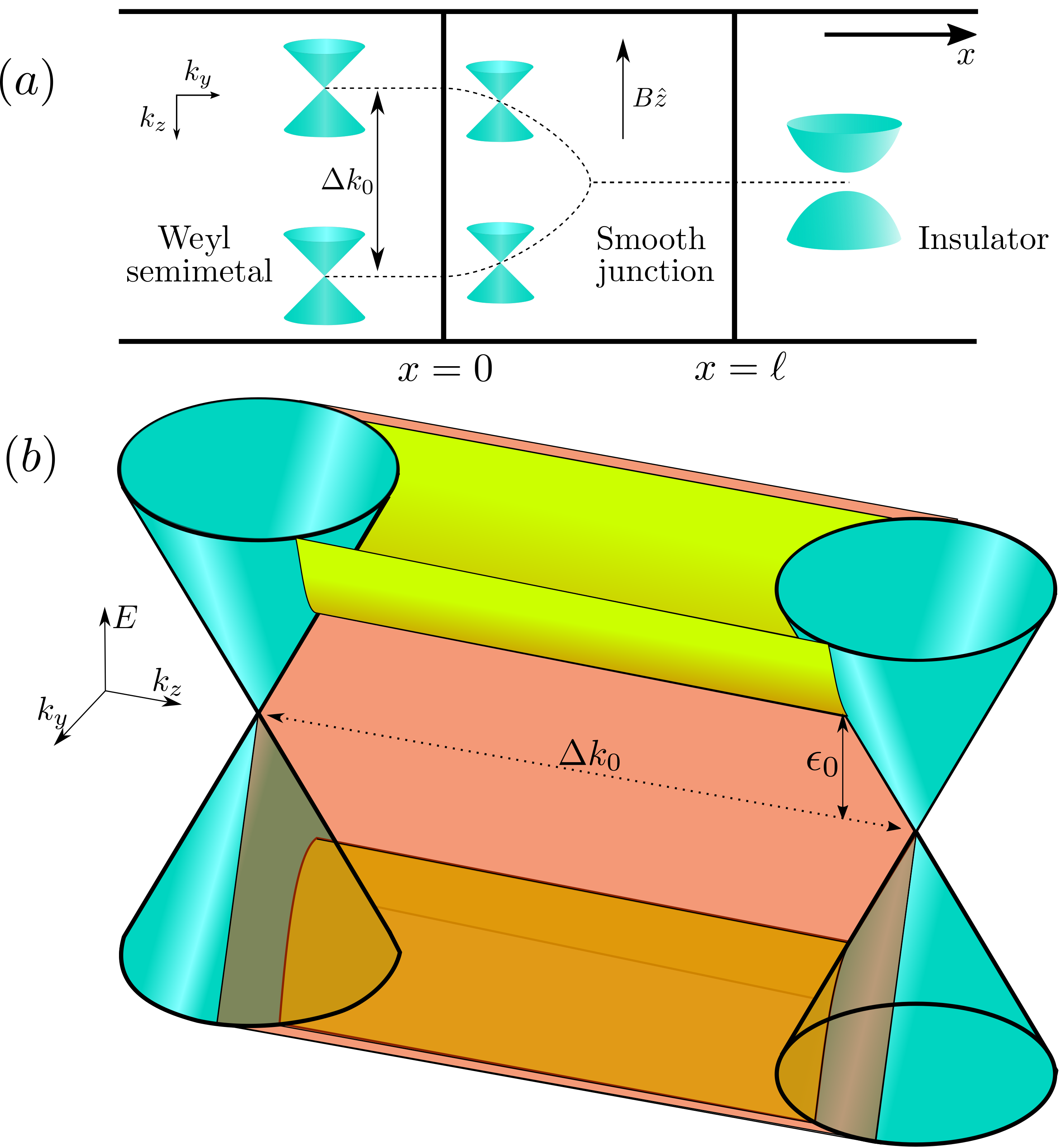}
    \caption{(a) Sketch of the smooth topological heterojunction. Inside the Weyl semimetal ($x<0$) the Weyl nodes are separated in the momentum space along the $k_z$ direction by $k_0=2\sqrt{2m\Delta}$. Inside the junction $(0<x<\ell)$, the Weyl nodes merge and gap out to finally form an insulating phase ($x>\ell$). (b) Sketch of the surface band structure. The Weyl nodes are split along the $k_z$ direction. The chiral Fermi arc (shown in pink) and the first VP state (shown in yellow) extend up to the bulk bands (shown in blue) along the $k_z$ axis.}
    \label{fig:diagrammaticrepresentation}
  \end{figure}
\end{center}

Weyl semimetals (WSMs) belong to a different class of topological materials where the bulk band gap vanishes. These are 3D systems with an even number of points in the Brillouin zone (BZ) called Weyl nodes where the conduction and the valence bands cross with a linear dispersion\cite{Turner2013,Rao2016,Armitage2018}. Over the past few years, they have become a subject of intense investigation, both theoretical and experimental\cite{Xu2015a,Xu2015b,Lv2015,Lu2015}. Even though they are gapless, WSMs host topologically protected surface states, which form an open arc connecting the projection of the Weyl nodes on the surface BZ. In order to appreciate this point consider the surface to be normal to the z-direction and the pair of Weyl nodes to be aligned along $k_z$ (Fig.\ref{fig:diagrammaticrepresentation}). At any fixed value of $k_z$ between the nodes, the 2D momentum space slices (spanned by $k_x$ and $k_y$) can be thought of as 2D Chern insulators which can host edge states\cite{Turner2013}. The Fermi arcs can then be thought of as collections of all these momenta points that have surface states. At low energies, the density of states of the bulk bands, which vanishes as $E^2$ with the energy $E$, is dominated by that of the surface Fermi arc with a nonzero constant density of states. We expect the surface states to play a significant role in experiments probing the transport and thermodynamic properties of WSMs. 

Recently, the effect of smooth boundaries on the band structure of WSMs has been studied\cite{Grushin2016,Araki2016,Tchoumakov2017}. Similar to the case described earlier, heterojunctions involving WSMs can also host massive Volkov-Pankratov (VP) bands if the junction is sufficiently smooth ($\ell\gg\xi$). Indeed, the spatial variation of the interface can be effectively described by a magnetic field so that the resulting spectrum is reminiscent of 2D dispersive (pseudo-)Landau bands, indexed by an integer number $n$. In this picture, the Fermi arc state can be thought of as the $n=0$ band which is the only surviving band in the limit of a sharp interface ($\ell\rightarrow 0$).

Magneto-optical studies of the bulk in similar systems have been previously performed\cite{Carbotte2013,Goerbig2019,Wang2017,Carbotte2014,Carbotte2018a,Thakur2018,Konye2018,Jiang2018,Duan2019,Carbotte2018b,Alisultanov2017,Behrends2017,Wang2017a,Wang2017b,Shao2016,Shao2015,Klier2015,Carbotte2015}. Magneto-optical spectroscopy remains a natural tool to explore details of the band structure as well as electromagnetic modes\cite{Long2018,Chen2019} in these systems. However, the surface states are much less studied from a magneto-optics point of view\cite{Shi2017}. This is precisely the aim of our present paper where we investigate the optical conductivity of the surface of a binodal WSM both in the presence and in the absence of a magnetic field parallel to the surface. The surface is modelled to be smooth enough to host multiple VP states. We demonstrate that not all optical transitions between the massive VP states are allowed, but the pseudo-Landau band like nature of these states gives rise to selection rules which have non-trivial effects on the response functions. We emphasize that a magnetic field is not necessary for the observation of these effects. However, the presence of the magnetic field modifies the selection rules by changing the velocity operators and subsequently the magneto-optical spectrum.

The article is organized as follows. In section II, we describe our model and obtain the VP states localized in an interface between a Weyl semimetal and an insulator. The density of states corresponding to these states is obtained and the conditions for their visibility are discussed. In section III, we obtain the absorptive components of the optical conductivity corresponding to these states. In section IV, we switch on a magnetic field parallel to the interface. This modifies the selection rules corresponding to optical transitions. Section V contains discussions and a summary of our results.

\section{VP states at the interface of a WSM-insulator heterojunction}

\begin{center}
  \begin{figure}
    \includegraphics[width=0.45\textwidth]{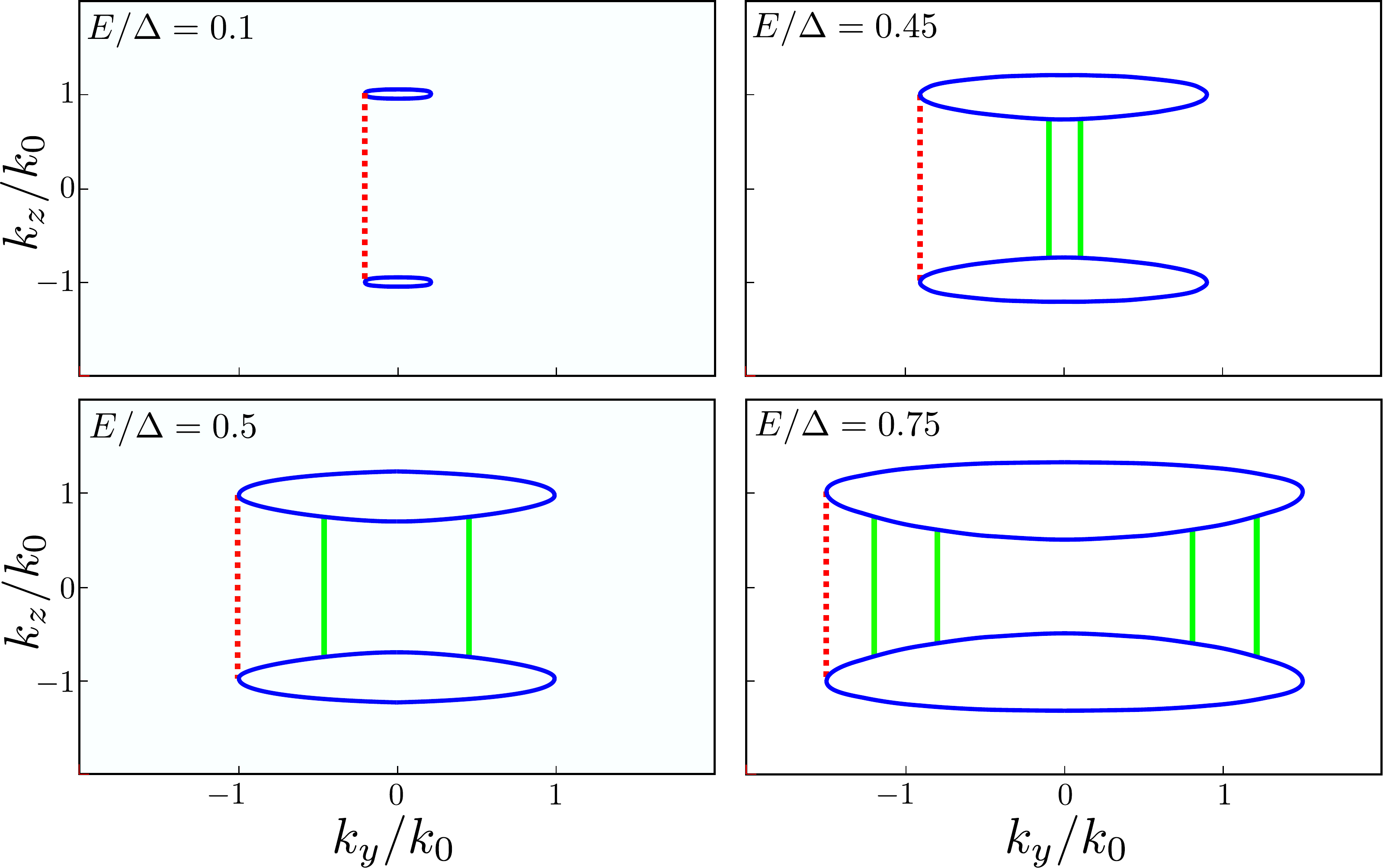}
      \caption{The states on the surface reciprocal space for a semi-infinite slab geometry for different energies. For low energies ($E=0.1\Delta$), only the Fermi arc (shown in dashed red line) is visible. As the energy is increased beyond $\epsilon_0(\approx 0.447\Delta$ for the chosen set of parameter values), the first VP band (in green lines) appears. For larger $E$, the second VP state appears. It is interesting to note that the boundary of the VP states along the $k_z$ axis do not change for changing $E$. The parameters used are $v_F=0.5$ eV $\angstrom$, $\Delta = 0.5$ eV, $\Delta'=\Delta$, $m=0.25$ eV$^{-1}\angstrom^{-2}$, $\ell = 20\angstrom$ and $\textbf{B}=0$.} 
      \label{fig:energycontourswithoutB}
    \end{figure}
\end{center}

We wish to construct the Hamiltonian for the interface of length $\ell$ between a Weyl semimetal bulk on the left and an insulator on the right. Our starting point is the Hamiltonian of the Weyl semimetal defined for $x<0$
\begin{align}
  \hat{H}_L = v_F(k_x\hat{\sigma}_x+k_y\hat{\sigma}_y) + \Big(k_z^2/2m-\Delta\Big)\hat{\sigma}_z
\end{align}
where we use a system of units with $\hbar=1$ and $\Delta > 0$.

This Hamiltonian is the simplest model which accounts for two Weyl nodes at $\textbf{k}_{\pm}=(0,0,\pm \sqrt{2m\Delta})$ with opposite topological charge. For $x>\ell$, one gets a trivial gapped phase for $\Delta\rightarrow -\Delta'< 0$. 

Now, the interface can be constructed by assuming that the parameter $\Delta$ varies linearly over the interface, smoothly connecting the two phases. Thus the Hamiltonian for $0\leq x \leq \ell$ can be written as\cite{Tchoumakov2017}
\begin{align}
  \hat{H}_S = v_F(k_x\hat{\sigma}_x+k_y\hat{\sigma}_y) + \Big(k_z^2/2m-\Delta+\frac{\Delta+\Delta'}{\ell}x\Big)\hat{\sigma}_z.
\end{align}
\begin{center}
  \begin{figure}
    \includegraphics[width=0.5\textwidth]{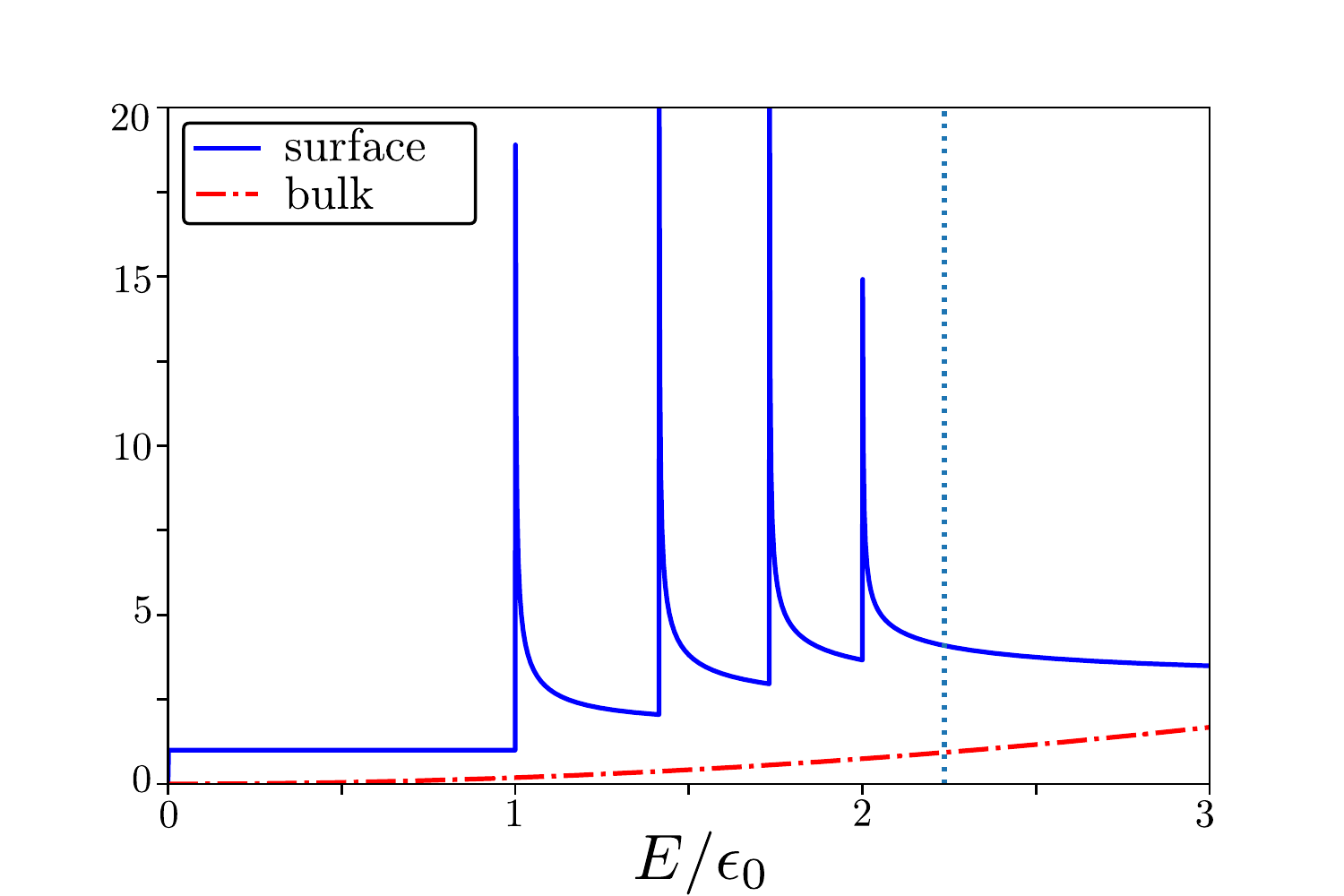}
    \caption{Comaprison of density of states of the VP states with that of the bulk. The bulk density of states is multiplied by $v_F/\Delta$ to make the two quantities dimensionally consistent. The dotted vertical line denotes the $E=\Delta$ point, beyond which no new VP state can appear. The parameters used are the same as in Fig. \ref{fig:energycontourswithoutB}. For these parameters, only four VP states can be seen.}
    \label{fig:VPdos}
  \end{figure}
\end{center}
We also consider a magnetic field along the direction of separation of the Weyl nodes and parallel to the interface $\textbf{B}=B\hat{z}$. In the Landau gauge, this is equivalent to substituting $k_y\rightarrow k_y+x/\ell^2_B$ where $\ell_B=1/\sqrt{eB}$ is the magnetic length. A sketch of the interface is shown in Fig. \ref{fig:diagrammaticrepresentation}(a). Finally, we perform a unitary rotation of the Hamiltonian in the $yz$ plane by an angle $\theta = \tan^{-1}(\ell_B^2/\ell_{Bp}^2)$. Here $\ell_{Bp}=1/\sqrt{eB_p}$ is a magnetic length associated with the pseudomagnetic field
\begin{align}
  B_p=(\Delta+\Delta')/ev_F\ell
  \label{eq:pseudomagnetic}
\end{align}
which arises from the variation of the gap at the interface. The purpose of this rotation is to gather all the terms linear in $x$ and $k_x$ in the off-diagonal components of the rotated Hamiltonian. 

In this rotated frame, the Hamiltonian for the surface can be written as\cite{Tchoumakov2017}
\begin{align}
  \hat{H}^{\theta}_{s} = e^{i\frac{\theta}{2}\hat{\sigma}_x}\hat{H}_se^{-i\frac{\theta}{2}\hat{\sigma}_x} = \left(\begin{array}{cc}
               M(\textbf{k}_{\parallel}) & \sqrt{2}v_F\hat{a}/\ell_S \\
                                  \sqrt{2}v_F\hat{a}^{\dagger}/\ell_S & -M(\textbf{k}_{\parallel})             \end{array}\right)
                                                                          \label{eq:HamiltonianVP}
\end{align}
where, $M(\textbf{k}_\parallel) = [B(k_z^2/2m - \Delta) - B_pv_Fk_y]/B_T$ in terms of the ``total'' magnetic field  $B_T = \sqrt{B^2+B_P^2}$.

Here, the ladder operators are given by
\begin{align}
  \hat{a}&=\frac{\ell_S}{\sqrt{2}}[k_x-i(x-\langle x\rangle)/\ell_S^2] \nonumber\\
  \hat{a}^\dagger&=\frac{\ell_S}{\sqrt{2}}[k_x+i(x-\langle x\rangle)/\ell_S^2],
\end{align}
and carry relevant information about the surface states. Indeed $\langle x\rangle = -[B_P(k_z^2/2m - \Delta) + Bv_Fk_y]/ev_FB_T^2$ determines the average position of the surface state, which has a Gaussian profile of a characteristic width $\ell_S=1/\sqrt{eB_T}$. 

The eigenstates of the surface Hamiltonian in Eq. (\ref{eq:HamiltonianVP}) are
\begin{align}
  |\psi_{n\lambda}\rangle &= \left(\begin{array}{c}
                                     u_{n\lambda}|n-1\rangle  \\ \lambda v_{n\lambda}|n\rangle \end{array}\right) \nonumber \\
  &= \frac{1}{\sqrt{2}}\left(\begin{array}{c} \left(1+\lambda\frac{M(\textbf{k}_\parallel)}{E_n}\right)^{1/2}|n-1\rangle  \\ \lambda\Big(1-\lambda\frac{M(\textbf{k}_\parallel)}{E_n}\Big)^{1/2}|n\rangle \end{array}\right).
\end{align}
Here, $|n\rangle$ corresponds to the eigenvectors of the number operator ($\hat{n}=\hat{a}^\dagger\hat{a}$) constructed from the ladder operators with eigenvalue $n$. The energy eigenvalues are
\begin{align}
  E_{n\lambda} = \lambda E_n = \lambda \sqrt{\frac{2v_F^2}{\ell_S^2}n + M(\textbf{k}_\parallel)^2}=\lambda\sqrt{\epsilon_1^2 n +M(\textbf{k}_\parallel)^2}
  \label{eq:VPspectrum}
\end{align}
where $\epsilon_1^2 = 2v_F^2/\ell_S^2$ and $\lambda=\pm 1$ characterizes the band index. While the $n=0$ gives rise to the chiral massless surface state $E_{n=0}=-M(\textbf{k}_\parallel)$, the states with $n\neq 0$ are the massive VP states with a gap $\epsilon_1\sqrt{n}$\footnote{A recent paper [Mahler et. al., Phys. Rev. X \textbf{9}, 031034(2019)] mentions that not only the massive but also the chiral surface states should be coined VP states. This is correct in the sense that both types of states arise from the same phenomena of gap inversion over the interface. However, in order not to give the impression that the n=0 VP state is something different from the extensively studied chiral massless states, we have decided to use the acronym VP only for the less common massive surface states.}. At this point, we also introduce $\epsilon_0^2=\epsilon_1^2(\textbf{B}=0)=2v_F(\Delta+\Delta')/\ell$. The two parameters, $\epsilon_1$ and $\epsilon_0$ parametrize the smoothness of the interface.
\begin{center}
  \begin{figure}
    \includegraphics[width=0.45\textwidth]{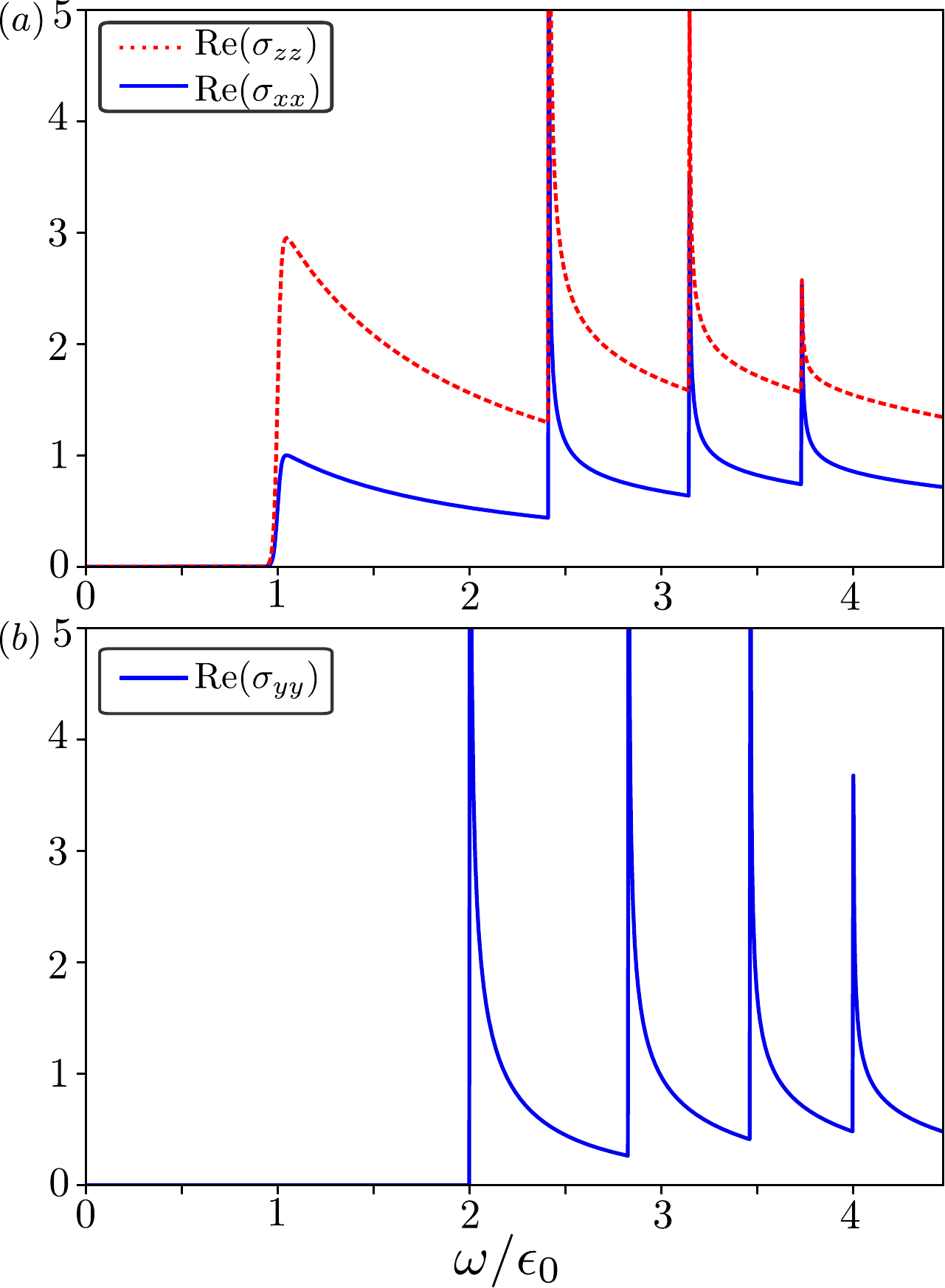}
    \caption{Longitudinal optical conductivities (in units of $e^2$) for $\mu=0$ along the $(a) x,z$ and $(b) y$ directions. All possible transitions involving the VP states are shown. In $(a)$, the poles are located at $\omega=\epsilon_0(\sqrt{n+1}+\sqrt{n})$ and in $(b)$, they are located at $\omega=2\sqrt{n}\epsilon_0$. The parameters used are the same as in Fig. (\ref{fig:energycontourswithoutB}).}
    \label{fig:opticalwithoutBnomu}
  \end{figure}
\end{center}
Notice that the spectrum in Eq. (\ref{eq:VPspectrum}) needs to be cut off at high energies roughly when the massive surface states are pushed into the inverted bulk gap ($\Delta$). By imposing the localization condition $(0\leq \langle x\rangle \leq \ell)$ ,i.e., the states must be situated within the interface of thickness $\ell$, it can be seen that just like the Fermi arc, the VP states exist only between the Weyl nodes in the momentum space. This is demonstrated in Fig. \ref{fig:diagrammaticrepresentation}(b) where the Fermi arc and the first VP state are shown in the limit $\textbf{B}=0$. In this limit, as the energy eigenvalues suggest, these bands disperse only along $k_y$. Fig. \ref{fig:energycontourswithoutB} shows how the VP states emerge as the chemical potential is increased beyond the corresponding gaps. We can make a rough estimate of the number of visible VP states ($n_1$) by analysing the length scales of the problem: $n_1 \lesssim \ell/\xi$ (where $\xi=v_F/\Delta$ is the bulk length scale).  
\begin{center}
  \begin{figure}[h]
    \includegraphics[width=0.45\textwidth]{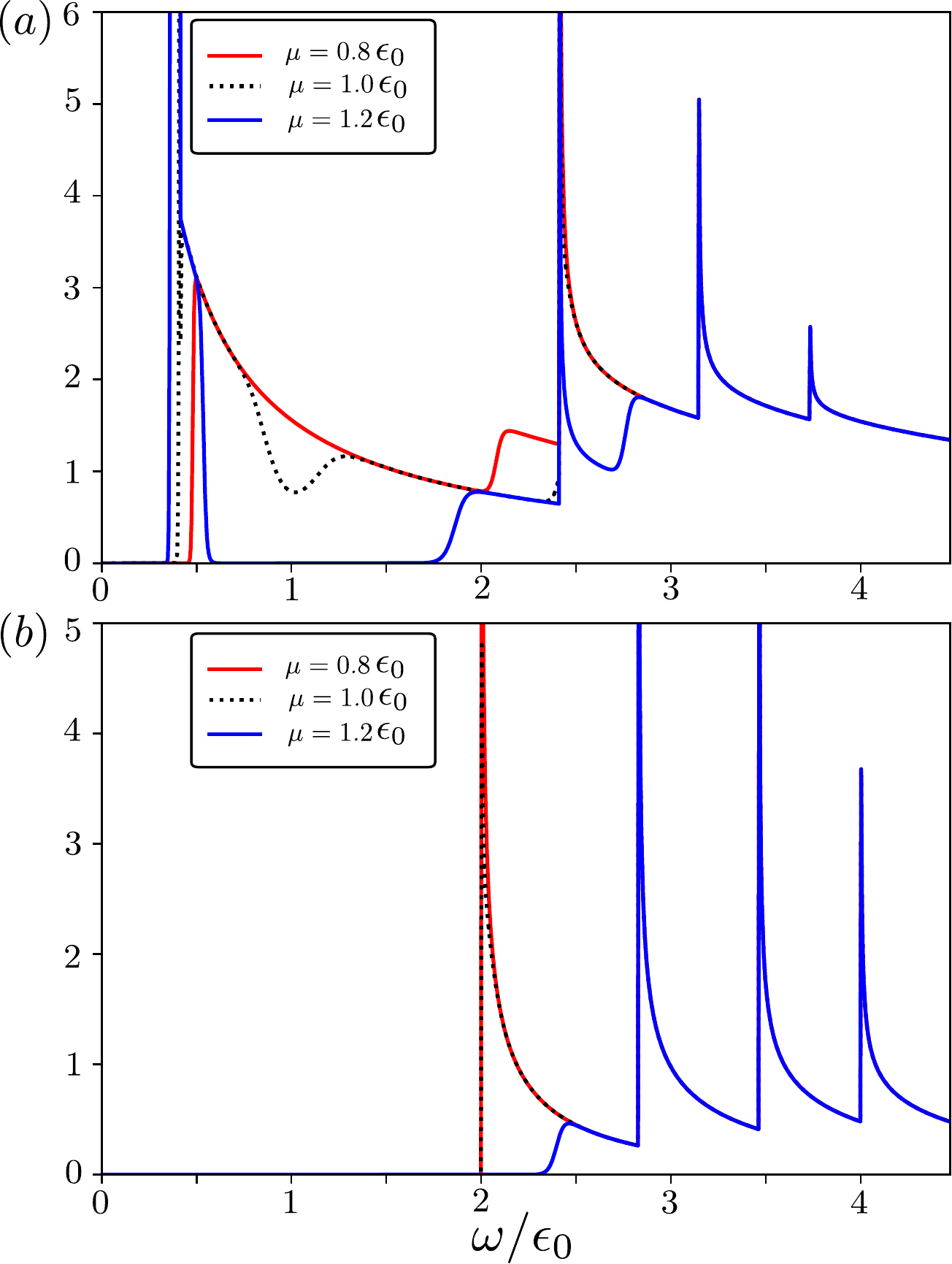}
    \caption{ $(a) \sigma_{zz}$ and $(b) \sigma_{yy}$  (in units of $e^2$) for different values of $\mu$. In $(a)$, as the chemical potential is increased, the absorption edge at $\omega=\epsilon_0$ splits into two. For larger chemical potential there is an overall redistribution of spectral weight and one can see a large accumulation for low frequencies (at $\omega\sim 0.5\epsilon_0$). In $(b)$, the plot remains unchanged until $\mu=\epsilon_0$ beyond which the conductivity vanishes for $\omega\leq \mu+\epsilon_0$ as the $n=1$ VP band starts getting occupied.}
    \label{fig:opticalwithmunoB}
  \end{figure}
\end{center}
It is to be noted here that the external magnetic field is not essential for the formation of these Landau bands. The pseudomagnetic field $B_p$, shown in Eq. (\ref{eq:pseudomagnetic}), is sufficient for their creation. The external field, however, changes the dispersion and also the optical properties as we show in section IV. Also, under the application of the magnetic field, the spacing between two band minima is enhanced by $\epsilon_1/\epsilon_0=[1-(\ell_S/\ell_B)^4]^{-1/4}$.

We first compute the density of states of the $n^{\text{th}}$ VP states for $\textbf{B}=0$ to get
\begin{align}
  \rho_n(E) = \frac{k_{zn}E}{2\pi^2v_F\sqrt{E^2-\epsilon_0^2 n}}\Theta(E-\sqrt{\epsilon_0^2 n})
\end{align}
where $\Theta(x)$ is the Heaviside step function [with $\Theta(x>0)=1$ and $\Theta(x<0)=0$] and $\pm k_{zn}$ denotes the points on the $z$ axis where the VP bands and the bulk bands intersect. This indicates the phase space available to the VP bands. Moreover, $k_{zn} = \sqrt{2m(\Delta - \epsilon_0\sqrt{ n})}$ and is independent of the energy as well as of the other transverse momentum $k_y$ when $\textbf{B}=0$. Details of the calculation are given in Appendix A.

The total density of states, $\rho(E) = \sum_{n}\rho_n(E)$ has poles at $E = \sqrt{n}\epsilon_0$ as shown in Fig. \ref{fig:VPdos}. Below $E=\epsilon_0$, only the chiral edge state is visible and the density of states is constant. The additional poles, whose nature resembles the van Hove singularities of 1D wires, appear when the energy is equal to the minima of the VP bands and strongly depend on the smoothness of the interface as well as on $\Delta$. Beyond the $n^{\text{th}}$ pole, $\rho_n(E)$ depends on energy as $\sim 1+\epsilon_0\sqrt{n}/E$ and has a long tail. 

In the limit $\ell\rightarrow 0$, $\epsilon_0$ diverges. Thus, in this limit only the chiral state is visible. In order to see the first nontrivial pole, we must have $\sqrt{n}\epsilon_0 < \Delta \Rightarrow \Delta\ell > 4v_F$ where we have taken $\Delta' = \Delta$. This means that for the visibility of the VP states either $\Delta$ has to be large or the surface has to be smooth ,i.e. $\ell$ has to be large. It is clear that the maximum level visible is given by $n_{\text{max}}=\left \lfloor{\ell/4\xi}\right \rfloor$ ,i.e. the integer part of $\ell/4\xi$. For example, in Fig. \ref{fig:VPdos}, we have four VP states for the parameters used. We point out that we obtain here in terms of energy considerations, a criterion for the number of visible surface states that is in line with that obtained from the localization condition. For higher levels, $\rho_{n}$ either vanishes or is imaginary. This is due to phase space restrictions. The VP states are present in the reciprocal space only between the Weyl nodes. For higher energies, the two disjoint bulk Fermi surfaces merge into one Fermi surface and hence do not allow the higher VP states to appear.

\section{Optical conductivity}

% \subsection{Selection rules and Kubo formula}

The frequency-dependent optical conductivity can be calculated by using the Kubo formula
\begin{align}
  \sigma_{ij}(\omega+i\eta)  &= i\sum_{m\lambda,n\lambda'}\int \frac{dk_ydk_z}{(2\pi)^2} \frac{f(E_{m\lambda})-f(E_{n\lambda'})}{\omega} \nonumber \\ &\hspace{2cm}\times\frac{(\hat{J}_i)_{m\lambda,n\lambda'}(\hat{J}_j)_{n\lambda',m\lambda}}{\omega+i\eta+E_{m\lambda}-E_{n\lambda'}} 
\end{align}
where $(\hat{J}_i)_{m\lambda,n\lambda'}= \langle \psi_{m\lambda}|\hat{J}_i|\psi_{n\lambda'}\rangle$ and $f(E)$ is the Fermi-Dirac distribution function. Here, the current operators are defined as $\hat{J}_i=\frac{\partial }{\partial k_i}\hat{H}^\theta_s$.

The real absorptive parts of the conductivity tensor are
\begin{widetext}
  \begin{align}\label{eq:sigma_xx}
    \text{Re}(\sigma_{xx}) &= 2e^2\frac{v_F}{\pi} \left( \frac{\sinh(\frac{\epsilon_0^2+\omega^2}{2\omega T})}{\cosh(\mu/T)+\cosh(\frac{\epsilon_0^2+\omega^2}{2\omega T})} - \frac{\sinh(\frac{\epsilon_0^2-\omega^2}{2\omega T})}{\cosh(\mu/T)+\cosh(\frac{\epsilon_0^2-\omega^2}{2\omega T})} \right) \sum_{n\geq 0} \frac{k_{z,n+1}|-\omega^2+(2n+1)\epsilon_0^2| }{\omega\sqrt{\omega^4+\epsilon_0^4-2\omega^2\epsilon_0^2(2n+1)}}\Theta'_{n}(\omega-\sqrt{n}\epsilon_0)
    \end{align}
  \begin{align}
    \text{Re}(\sigma_{yy}) &= 2\frac{e^2v_F\epsilon_0^2}{\pi} \frac{\sinh(\omega/2T)}{\cosh(\mu/T)+\cosh(\omega/2T)} \sum_{n\geq 0} \frac{nk_{zn}}{(\omega/2)^2\sqrt{(\omega/2-\sqrt{n}\epsilon_0)}\sqrt{(\omega/2+\sqrt{n}\epsilon_0)}} \Theta(\omega-2\sqrt{n}\epsilon_0)
  \end{align}
  \begin{align}\label{eq:sigma_zz}
    \text{Re}(\sigma_{zz}) &=  \frac{2e^2}{3m^2v_F\pi} \left( \frac{\sinh(\frac{\epsilon_0^2+\omega^2}{2\omega T})}{\cosh(\mu/T)+\cosh(\frac{\epsilon_0^2+\omega^2}{2\omega T})} - \frac{\sinh(\frac{\epsilon_0^2-\omega^2}{2\omega T})}{\cosh(\mu/T)+\cosh(\frac{\epsilon_0^2-\omega^2}{2\omega T})} \right) \sum_{n\geq 0} \frac{k^3_{z,n+1}|-\omega^2+\epsilon_0^2(2n+1)| }{\omega\sqrt{\omega^4+\epsilon_0^4-2\omega^2\epsilon_0^2(2n+1)}}\Theta'_{n}(\omega-\sqrt{n}\epsilon_0)
\end{align}
\end{widetext}
where $T = 0.01\epsilon_0$ is the temperature in a system of units with the Boltzmann constant $k_B=1$ and $\Theta'_n(x)=\Theta(x)$ is the Heaviside step function for $n\geq 1$ and $\Theta'_0(x)=1$. The outline of the derivation is given in Appendix B. Eqs. (\ref{eq:sigma_xx})-(\ref{eq:sigma_zz}) form the first set of major results of this article.

In Fig. \ref{fig:opticalwithoutBnomu}, we show the variation of the longitudinal conductivity as a function of $\omega$ for $\textbf{B}=0$ and $\mu=0$. The allowed transitions for $\sigma_{xx}$ and $\sigma_{zz}$ are $n\rightarrow n \pm 1$, regardless of $\lambda$. As a consequence, the poles for Fig. \ref{fig:opticalwithoutBnomu}(a) are located at $\omega = (\sqrt{n}+\sqrt{n+1})\epsilon_0$ for $n= 1,2,...,n_{\text{max}}-1$. Just like the case of the density of states, the peak spacing depends on $\epsilon_0 \propto 1/\sqrt{\ell}$ and hence depends on the smoothness of the interface as well as the inverted band gap $\Delta$. However, there is no actual pole for the $n=0$ to $n=\pm 1$ transition. The edge at $\omega \approx \epsilon_0$ appears due to vanishing Fermi factors, unlike the other poles, where the denominators actually vanish. This can be attributed to the flat density of states of the chiral Fermi arc, which prohibits the joint density of states from diverging.

The conductivity $\sigma_{yy}$ couples the bands with same energy index, i.e., only $n\rightarrow n$ transitions are allowed with different $\lambda$. So, the real part of $\sigma_{yy}$ in Fig. \ref{fig:opticalwithoutBnomu}(b) has poles at $\omega=2\sqrt{n}\epsilon_0$ for $n = 1,2,..., n_{\text{max}}$. Such a transition that preserves momentum is not possible for the chiral Fermi arc state and hence $\sigma_{yy}=0$ below $\omega=2\epsilon_0$.

Lastly, the matrix elements of the current operators $J_x$ and $J_z$ involve selection rules which are incompatible with those of $J_y$, resulting in $\sigma_{xy}=\sigma_{yz}=0$. Also, $\sigma_{xz}$ is an odd integral under $k_z\rightarrow -k_z$ and hence, vanishes when summed over momentum space.

\begin{center}
  \begin{figure}
    \includegraphics[width=0.45\textwidth]{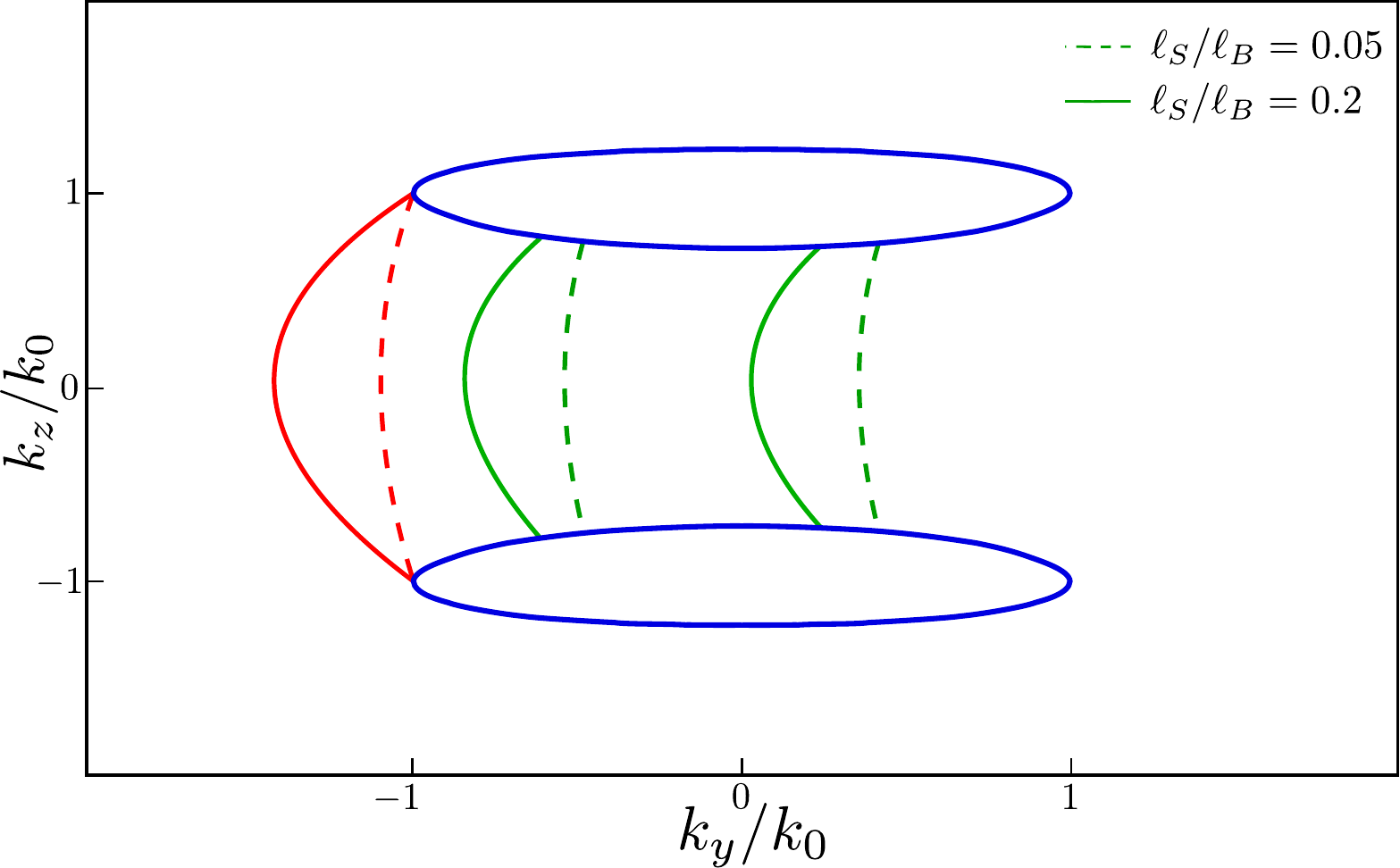}
    \caption{Constant energy contours for $E=0.5\Delta$ with two different values of magnetic field. The continuous lines denote the contours for $\ell_S/\ell_B=0.2$ and the dashed lines are for $\ell_S/\ell_B=0.05$. The two red lines on the left denotes the $n=0$ Fermi arc states for the different values of the magnetic field whereas the green lines depict the $n=1$ VP states. Interestingly, the change in the length of the VP states along the $k_z$ axis is very small for small magnetic fields and is neglected in the optical conductivity calculations. }
    \label{fig:energycontourswithB}
  \end{figure}
\end{center}

In Fig. \ref{fig:opticalwithmunoB} we plot the frequency dependence of $\sigma_{zz}$ and $\sigma_{yy}$ as the chemical potential is increased from zero. The $\omega=\epsilon_0$ absorption edge in $\sigma_{zz}$ (shown in Fig. \ref{fig:opticalwithmunoB}(a)) immediately splits into two distinct edges at $\omega_\pm\approx\sqrt{\mu^2+\epsilon_0^2}\pm \mu$. These features appear because with the introduction of the chemical potential, the transitions from $(-)1\rightarrow 0$ and $0\rightarrow (+)1$ bands require different energies. Consequently, the spectral weight is redistributed. As the chemical potential is further increased beyond the $n=1$ band minimum, we find that for $\sigma_{zz}$, the peak due to $1\rightarrow 2$ transitions is diminished. Instead we get an accumulation of spectral weight for low frequencies ($\omega \sim 0.5\epsilon_1 $). This is due to the transitions involving the Fermi arc and will be present for larger chemical potentials also as this state envelopes the $n\geq 1$ VP states at all energies. The qualitative behaviour of $\sigma_{xx}$ is similar and is not shown here.

On the other hand $\sigma_{yy}$, as explained earlier, does not involve the Fermi arc and has contributions from $n\rightarrow -n$ transitions. As a result, $\sigma_{yy}$ remains unaffected by the increasing chemical potential as long as $\mu < \epsilon_0$. However, beyond this value, the number of possible interband transitions is reduced as the $n=(+)1$ VP band starts to get occupied. Consequently, the finite frequency conductivity vanishes below $\omega = \mu + \epsilon_0$ as shown in Fig. \ref{fig:opticalwithmunoB}(b). This vanished spectral weight will be balanced by an increased Drude peak.

%For both $\sigma_{xx}$ and $\sigma_{zz}$, the peaks involving the VP bands with $n \geq 1$ are unaffected as long as $\mu < \epsilon_1$. In Fig. \ref{fig:opticalwithBandmu}(c) and \ref{fig:opticalwithBandmu}(d) we plot the optical conductivities when the chemical potential ($\mu=1.2\epsilon_1$) is greater than the first VP band minimum (at $\epsilon_1$).  For $\sigma_{yy}$ also we see the same low frequency features. However, this feature was absent in Fig. \ref{fig:opticalwithmunoB}(b) where we plot the frequency dependence of $\sigma_{yy}$ for various values of chemical potentials in the absence of $\textbf{B}$. This is because when $\textbf{B}=0$, the selection rules prohibited processes involving the $n=0$ Fermi arc state.

\section{Optical conductivity with magnetic field}

\begin{center}
  \begin{figure}[h]
    \includegraphics[width=0.45\textwidth]{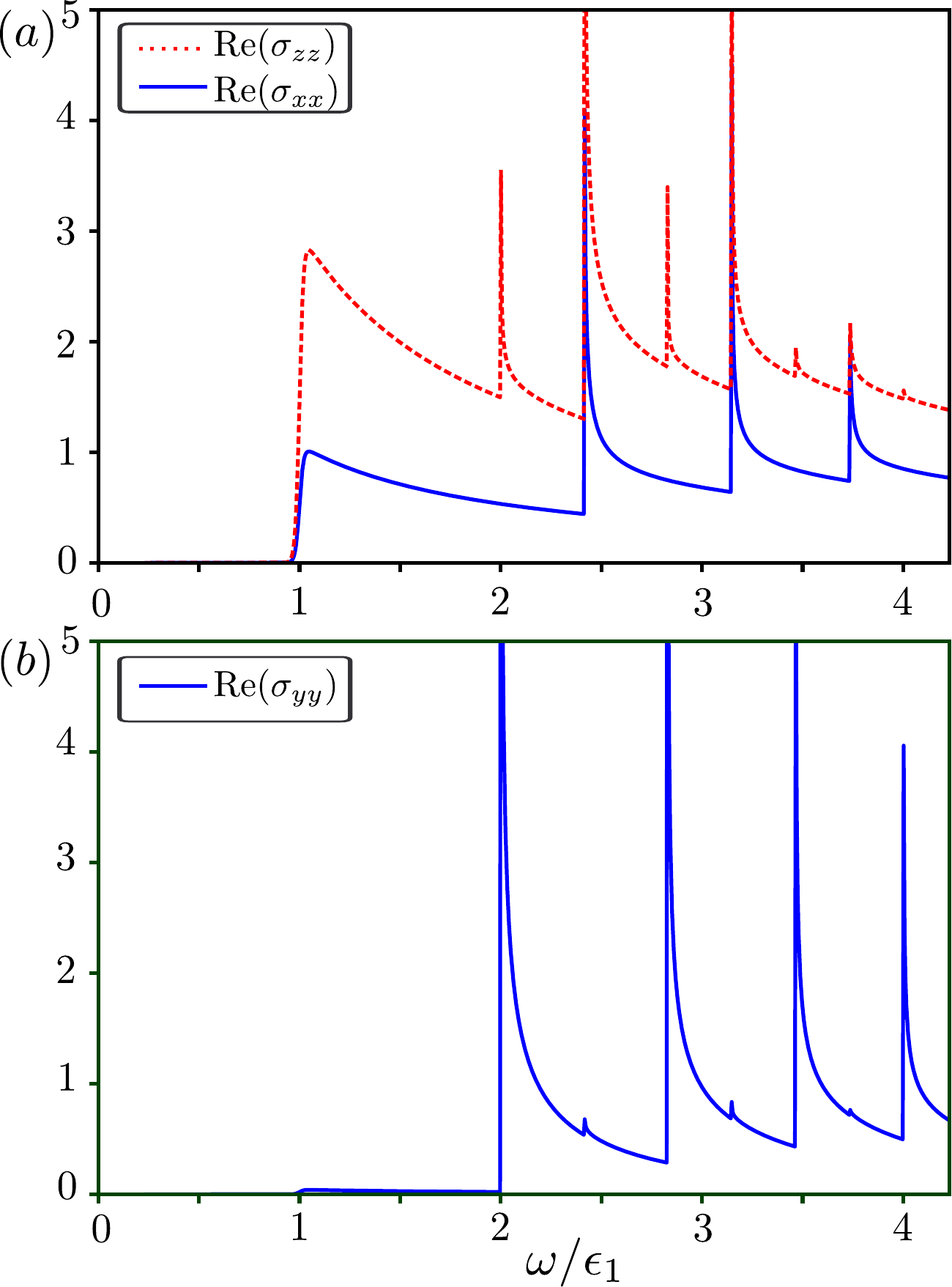}
    \caption{Longitudinal optical conductivities (in units of $e^2$) for $\mu=0$, $\ell_S/\ell_B=0.2$ $(a)$ along x and z and $(b)$ along y. In $(a)$, the real part of $\sigma_{zz}$ (red dashed line) has poles at both $\omega=\epsilon_1(\sqrt{n+1}+\sqrt{n})$ and $\omega=2\sqrt{n}\epsilon_1$. The real part of $\sigma_{xx}$ (blue continous line) remains unchanged apart from an overall scaling along the x axis due to the change of the effective magnetic field. In panel $(b)$, new poles at $\omega=\epsilon_1(\sqrt{n+1}+\sqrt{n})$, though very small, can be seen. Also, the effect of the absorption edge due to transitions from the Fermi arc can be seen below $\omega=2\epsilon_1$. This was prohibited in the $\textbf{B}=0$ limit. }
    \label{fig:opticalwithBnomu}
  \end{figure}
\end{center}

Application of the magnetic field changes the constant energy contours as can be seen in Fig. \ref{fig:energycontourswithB}. As a consequence, the $J_y$ and $J_z$ operators are mixed. This results in new allowed transitions in $\sigma_{yy}$ and $\sigma_{zz}$ , as can be seen in Fig. \ref{fig:opticalwithBnomu}. 

The nonzero components of the real absorptive part of the conductivity tensor in this case are given by
\begin{widetext}
\begin{align}\label{eq:sigma_xx_B}
  \text{Re}(\sigma_{xx})= &\frac{B_T}{B_P}\frac{2e^2v_F}{\pi} F_1(\omega) \sum_{n} \frac{k_{z,n+1}|\omega^2-\epsilon_1^2(2n+1)|}{\omega\sqrt{\omega^4+\epsilon_1^4-2\omega^2\epsilon_1^2(2n+1)}} \Theta'_{n}(\omega-\sqrt{n}\epsilon_1)
\end{align}
\begin{align}
  \text{Re}(\sigma_{yy}) = &\frac{B_P}{B_T}\frac{2e^2v_F\epsilon_1^2}{\pi} F_2(\omega) \sum_{n} \frac{nk_{z,n+1}}{(\omega/2)^2\sqrt{\omega^2/4-\epsilon_1^2 n} } \Theta(\omega-2\sqrt{n}\epsilon_1) \nonumber \\ + &\frac{B^2}{B_PB_T} \frac{2e^2v_F}{\pi}  F_1(\omega) \sum_{n} \frac{k_{z,n+1}|\omega^2-\epsilon_1^2(2n+1)| }{\omega\sqrt{\omega^4+\epsilon_1^4-2\omega^2\epsilon_1^2(2n+1)}} \Theta'_{n}(\omega-\sqrt{ n}\epsilon_1)
\end{align}
\begin{align}
  \text{Re}(\sigma_{zz}) = &\frac{B_P}{B_T} \frac{2e^2}{3m^2\pi v_F}  F_1(\omega) \sum_{n} \frac{k^3_{z,n+1}|\omega^2-\epsilon_1^2(2n+1)| }{\omega\sqrt{\omega^4+\epsilon_1^4-2\omega^2\epsilon_1^2(2n+1)}} \Theta'_{n}(\omega-\sqrt{n}\epsilon_1) \nonumber \\ + &\frac{B^2}{B_TB_P}\frac{2e^2\epsilon_1^2}{3m^2\pi v_F} F_2(\omega) \sum_{n} \frac{nk^3_{z,n+1}}{(\omega/2)^2\sqrt{\omega^2/4-\epsilon_1^2 n}} \Theta(\omega-2\sqrt{ n}\epsilon_1)
\end{align}
  \begin{align}\label{eq:sigma_xy_B}
  \text{Im}(\sigma_{xy}) = &- \frac{B}{B_P}\frac{2e^2v_F}{\pi}F_3(\omega)\sum_{n} \frac{k_{z,n+1}|\omega^2-\epsilon_1^2(2n+1)|}{\omega\sqrt{\omega^4+\epsilon_1^4-2\omega^2\epsilon_1^2(2n+1)}} \Theta'_{n}(\omega-\sqrt{n}\epsilon_1)
\end{align}
where $F_1(\omega) =\frac{\sinh(\frac{\epsilon_1^2+\omega^2}{2\omega T})}{\cosh(\mu/T)+\cosh(\frac{\epsilon_1^2+\omega^2}{2\omega T})}-\frac{\sinh(\frac{\epsilon_1^2-\omega^2}{2\omega T})}{\cosh(\mu/T)+\cosh(\frac{\epsilon_1^2-\omega^2}{2\omega T})} $, $F_2(\omega)=\frac{\sinh(\omega/2T)}{\cosh(\mu/T)+\cosh(\omega/2T)}$ and $F_3(\omega) = \frac{e^{\mu/T}+\cosh(\frac{\omega^2+\epsilon_1^2}{2\omega T})}{\cosh(\mu/T)+\cosh(\frac{\omega^2+\epsilon_1^2}{2\omega T})}- \frac{e^{\mu/T}+\cosh(\frac{\omega^2-\epsilon_1^2}{2\omega T})}{\cosh(\mu/T)+\cosh(\frac{\omega^2-\epsilon_1^2}{2\omega T})}$.
\end{widetext}

Also, we have assumed that even though there is a small magnetic field, $\int dk_z\approx 2k_{zn}$.

In the absence of the magnetic field, $B_T=B_P$ and $\epsilon_1=\epsilon_0$. As is expected, in this limit, we reproduce the results from the previous section. Eqs. (\ref{eq:sigma_xx_B})-(\ref{eq:sigma_xy_B}) are the second set of major results of this work.

% \begin{center}
\twocolumngrid
  \begin{figure}
    \includegraphics[width=0.45\textwidth]{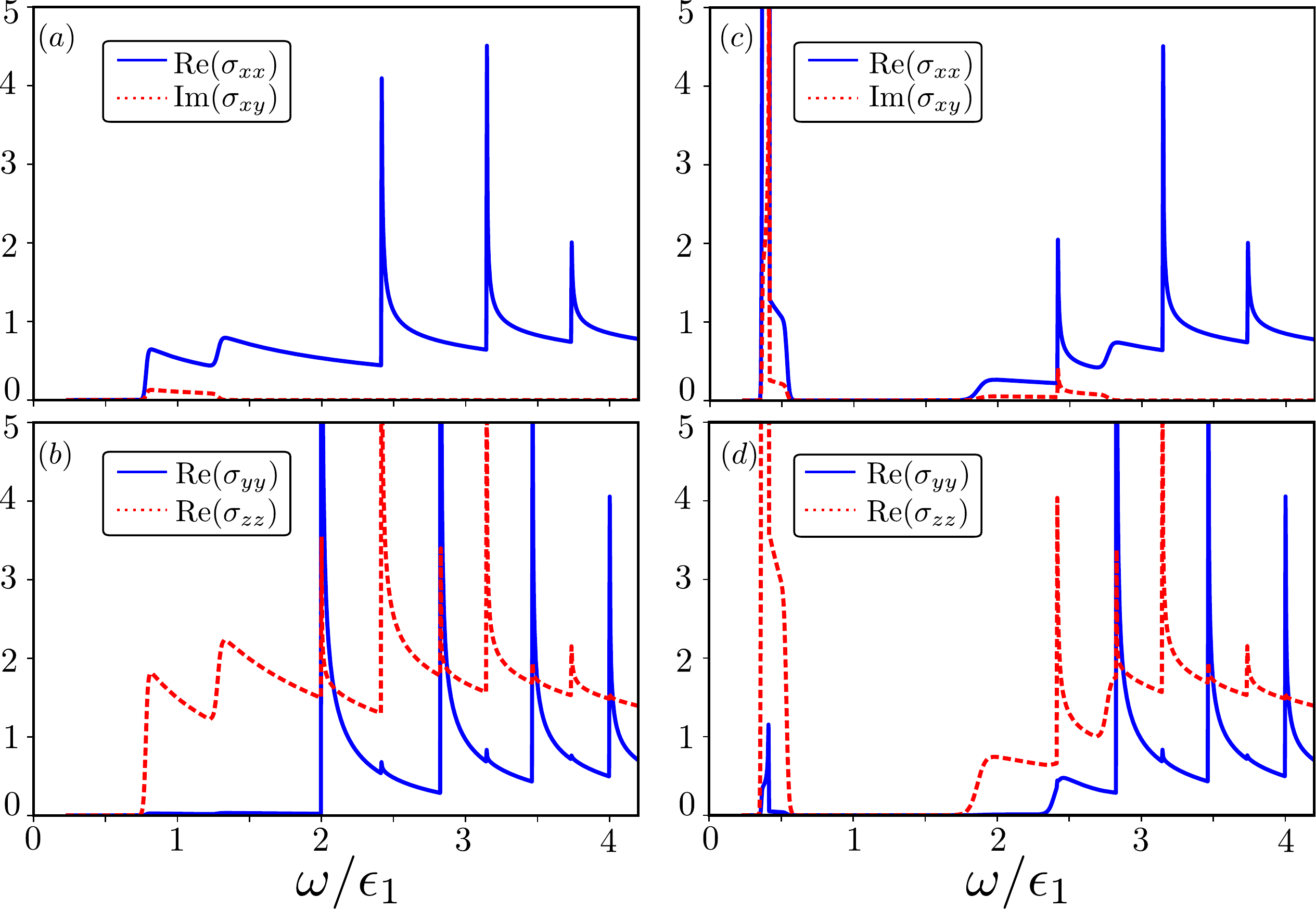}
    \caption{Optical conductivities (in units of $e^2$) by changing $\mu$ for $\ell_S/\ell_B=0.2$. We have used $\mu=0.25\epsilon_1$ in panels $(a)$ and $(b)$ and $\mu=1.2\epsilon_1$ in panels $(c)$ and $(d)$.}
    \label{fig:opticalwithBandmu}
  \end{figure}
%\end{center}

Due to the external field, the poles shift because of the change in the strength of the effective magnetic field $B_T=\sqrt{B_P^2+B^2}$ by a factor of $\epsilon_1/\epsilon_0$. This leads to an overall rescaling of the $x$ axis in Fig. \ref{fig:opticalwithBnomu}. However, it is to be noted that for the parameters used in this work, this scaling is almost equal to unity. The shifting of poles is demonstrated in Appendix C for a stronger magnetic field. The selection rules in $\sigma_{xx}$ remain unaffacted. This is because the application of the magnetic field does not change the current operator $J_x$. As a result, the overall nature of $\sigma_{xx}$ in Fig. \ref{fig:opticalwithBnomu}(a) remains unchanged from the previous section.

The real part of $\sigma_{zz}$, shown as the dotted line in Fig. \ref{fig:opticalwithBnomu}(a), has new poles at $\omega=2\sqrt{n}\epsilon_1$. This implies $n\rightarrow n$ (with different $\lambda$) transitions which were not allowed without the magnetic field. Indeed the magnetic field mixes the $z-$ and the $y-$ components of the conductivity, which leads to the additional peaks, in analogy to the situation encountered in topological insulators with smooth interfaces\cite{Xin2019}. These new poles exist along with the ones at $\omega=\epsilon_1(\sqrt{n+1}+\sqrt{n})$.

Similarly, in Fig. \ref{fig:opticalwithBnomu}(b), the new poles of $\sigma_{yy}$ at $\omega=\epsilon_1(\sqrt{n+1}+\sqrt{n})$ are visible, again indicating new selection rules. There are very small features below $\omega=2\epsilon_1$ which was previously prohibited in the $\textbf{B}=0$ case.

When $\mu=0$, $F_3(\omega)$ also vanishes, but, for finite chemical potential, $F_3(\omega)$ and subsequently the imaginary part of $\sigma_{xy}$ are nonzero. This is a consequence of time-reversal symmetry of the model being explicitly broken by the application of the external magnetic field. The other off-diagonal componenets, $\sigma_{yz}$ and $\sigma_{xz}$, are still odd under $k_z\rightarrow -k_z$ and vanish just like in the case when $\textbf{B}=0$. This is shown in Figs. \ref{fig:opticalwithBandmu}(a) and \ref{fig:opticalwithBandmu}(c) where Im($\sigma_{xy}$) is plotted for $\mu=0.25\epsilon_1$ and $\mu=1.2\epsilon_1$ respectively. The behaviour of $\sigma_{xx}$ remains unchanged from the case of no magnetic field. The splitting of the absorption edge, subsequent reduction of higher peaks and new low frequency peaks are also present in this case and can be seen in Figs. \ref{fig:opticalwithBandmu}(a) and \ref{fig:opticalwithBandmu}(c). In the limit where $B_P=0$ and $B_T=B$, one can show that the imaginary part of $\sigma_{xy}$ is exactly equal to the real part of $\sigma_{xx}$ for low frequencies, but it is not so here due to the additional pseudomagnetic field. Perhaps the most striking difference can be seen for $\sigma_{yy}$ where new low frequency poles can be seen for large $\mu$, which was previously prohibited by the selection rules in the $\textbf{B}=0$ case. This demonstrates the mixing of the velocity operators.

\section{Discussion and conclusion}

In conclusion, we have calculated the density of states and the magneto-optical conductivity of a smooth topological heterojunction between a WSM and an insulator. We find signatures of massive VP states in both these observables. These signatures explicitly depend on the inverted band gap ($\Delta$) of the WSM as well as on the smoothness of the interface ($\ell$). For the conductivity calculation, we find selection rules governing the available optical transitions. Even without the external magnetic field, the nature of $\sigma_{xx}$ resembles that of the bulk with a magnetic field\cite{Carbotte2013}. However, unlike the bulk case, where the magneto-optical conductivity peaks were seen on a linear background as a function of $\omega$\cite{Carbotte2013}, here no such background is seen. This is because of phase space constraints. The phase space availability of VP states is heavily suppressed as the two Fermi surfaces grow closer for higher energies. When a magnetic field parallel to the interface is turned on, the velocity components are mixed. As a result, the selection rules are also modified, leading to new peaks in the conductivities.

We acknowledge financial support from Agence Nationale de Recherche under grant no. ANR-17-CE30-0023 ``Dirac 3D".

\bibliography{References}

\appendix

\section{Details of the derivation of the DOS}
$\frac{\partial E_n}{\partial k_y} = \frac{v_F^2k_y}{\sqrt{v_F^2k_y^2 + \epsilon_0^2 n}} = \frac{v_F\sqrt{E_n^2 - \epsilon_0^2 n}}{E_n}$.

Hence,
\begin{align}
  \rho_n(E) &= \frac{1}{4\pi^2}\int \frac{dk_z}{|\partial E_n/\partial k_y|} \nonumber\\
          &=\frac{E_n}{4\pi^2v_F\sqrt{E_n^2-\epsilon_0^2 n}}\int dk_z \nonumber \\
          &=\frac{E_n}{4\pi^2v_F\sqrt{E_n^2-\epsilon_0^2 n}}\times 2k_{zn}
\end{align}
where $\pm k_{zn}$ are the points where the VP states intersect the bulk Fermi surface.

Now, the locus of the bulk Fermi surface on the $k_y,k_z$ plane is given by
\begin{align}
  &E^2 = v_F^2k_y^2 + (k_z^2/2m-\Delta)^2 \nonumber \\
  \Rightarrow & k_z = \pm\sqrt{2m(\Delta\pm\sqrt{E^2-v_F^2k_y^2})} .
\end{align}

For this Fermi surface to intersect with the VP states, $E = E_n = \sqrt{v_F^2k_y^2 + \epsilon_0^2 n} \Rightarrow k_{zn} = \sqrt{2m(\Delta - \sqrt{n}\epsilon_0)}$. This is independent of the energy as well as the transverse momentum $k_y$.

\begin{center}
  \begin{figure}[t]
    \includegraphics[width=0.45\textwidth]{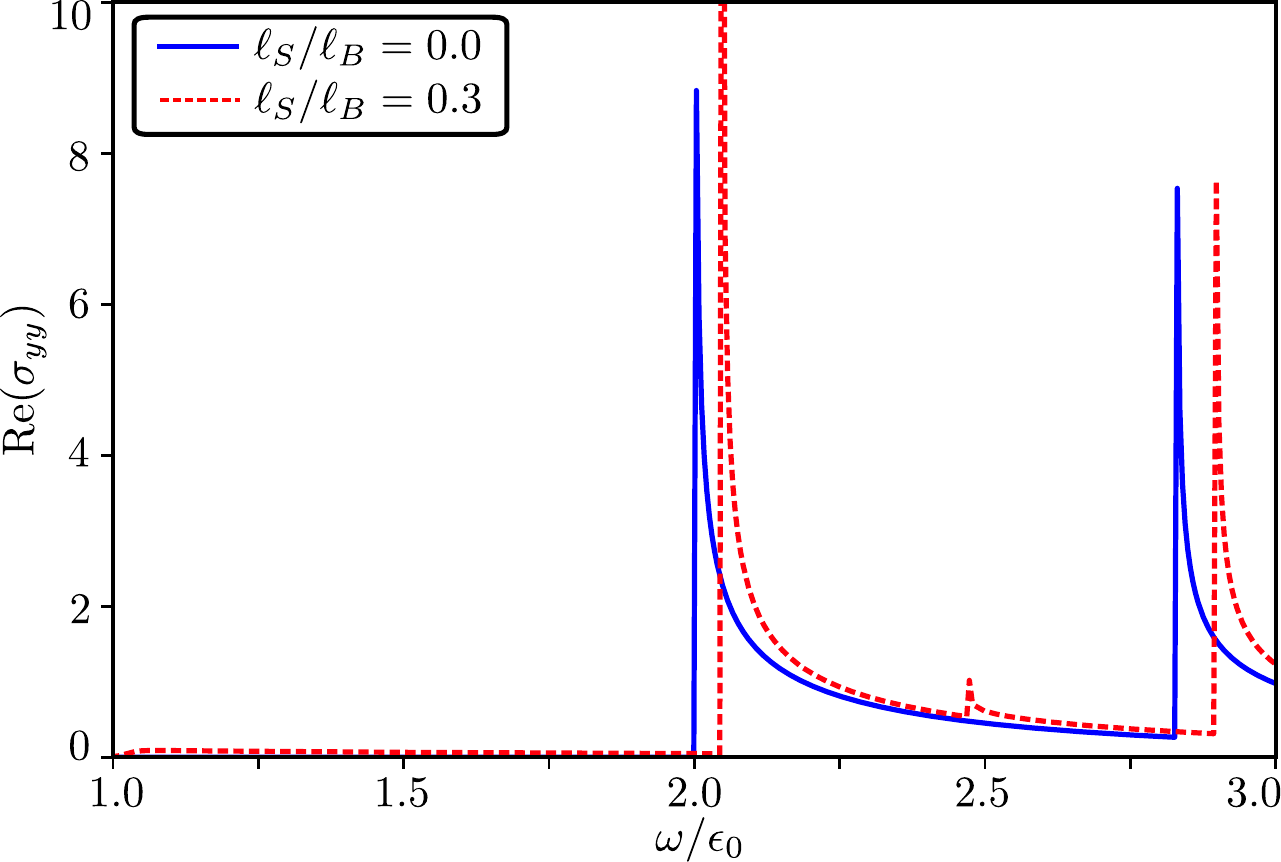}
    \caption{Comparison of the poles of $\sigma_{yy}$ with and without the magnetic field.}
    \label{fig:poleshift}
  \end{figure}
\end{center}

\section{Details of the Optical Conductivity calculation}
The current operators are:
\begin{align}
  \hat{J}_{x} &= e\left(\begin{array}{cc}
                         0 & v_F \\
                         v_F & 0             \end{array}\right) = ev_F\sigma_x\\
  \hat{J}_{y} &= e\left(\begin{array}{cc}
                         -\frac{B_P}{B_T}v_F & -i\frac{B}{B_T}v_F \\
                         i\frac{B}{B_T}v_F & \frac{B_P}{B_T}v_F             \end{array}\right) = -e\frac{B_P}{B_T}v_F\sigma_z + e\frac{B}{B_T}v_F\sigma_y \\
  \hat{J}_{z} &= e\left(\begin{array}{cc}
                         \frac{B}{B_T} \frac{k_z}{m} & -i\frac{B_p}{B_T} \frac{k_z}{m} \\
                         i\frac{B_p}{B_T} \frac{k_z}{m} & -\frac{B}{B_T}\frac{k_z}{m}             \end{array}\right) = e\frac{B}{B_T}\frac{k_z}{m}\sigma_z + e\frac{B_P}{B_T}\frac{k_z}{m}\sigma_y
\end{align}

Thus, the matrix elements are
\begin{widetext}
\begin{align}
  (J_x)_{m\lambda,n\lambda'}=& ev_F[\lambda u^*_{m\lambda}v_{n\lambda'}\delta_{m,n+1}+\lambda'u_{n\lambda'}v^*_{m\lambda}\delta_{m,n-1}] \\
  (J_y)_{m\lambda,n\lambda'}=& -ev_F\frac{B_P}{B_T}[\lambda\lambda' u^*_{n\lambda}u_{n\lambda'}-v_{n\lambda'}v^*_{n\lambda}]\delta_{m,n} -ev_F\frac{B}{B_T}[-i\lambda u^*_{m\lambda}v_{n\lambda'}\delta_{m,n+1}+i\lambda'u_{n\lambda'}v^*_{m\lambda}\delta_{m,n-1}] \\
  ( J_z)_{m\lambda,n\lambda'}=& e\frac{B}{B_T}\frac{k_z}{m}[\lambda\lambda' u^*_{n\lambda}u_{n\lambda'}-v_{n\lambda'}v^*_{n\lambda}]\delta_{m,n} +e\frac{B_P}{B_T}\frac{k_z}{m}[-i\lambda u^*_{m\lambda}v_{n\lambda'}\delta_{m,n+1}+i\lambda'u_{n\lambda'}v^*_{m\lambda}\delta_{m,n-1}] 
\end{align}

Lastly, since the current operator $J_z$ is odd under $k_z\rightarrow -k_z$, we obtain $\sigma_{xz} = \sigma_{yz} = 0$.

Here, we outline the calculation for the $\textbf{B}=0$ limit.

  \begin{align}
    \sigma_{xx}(\omega+i\eta) = i\sum_{m\lambda,n\lambda'}\int \frac{dk_ydk_z}{(2\pi)^2}&\frac{f(\lambda E_m) - f(\lambda' E_n)}{\omega}\frac{(J_x)_{m\lambda,n\lambda'}(J_x)_{n\lambda',m\lambda}}{\omega+i\eta+\lambda E_m-\lambda'E_n}
  \end{align}
  Looking only at the real absorptive part of the conductivity, 
  
  \begin{align}
    \text{Re}\Big(\sigma_{xx}(\omega)\Big) = -e^2v_F^2 \sum_{n}\frac{k_{z,n+1}}{\omega\pi}\int dk_y\Big[ & \Big(f(- E_{n+1}) - f(- E_n)-f(E_{n+1}) + f(E_n)\Big) (1-\frac{ M^2}{E_nE_{n+1}}) \delta(\omega+ E_{n}- E_{n+1})   \nonumber \\
                                            +& \Big(f(- E_{n+1}) - f( E_n)- f(E_{n+1}) + f(- E_n)\Big) (1+\frac{ M^2}{E_nE_{n+1}}) \delta(\omega- E_{n+1}-E_n) \Big]
  \end{align}

At this point, we change the integration variable from $k_y$ to $M(\textbf{k}_\parallel)$. The delta functions contribute when $\omega=E_{n+1}\pm E_n$. These conditions are satisfied when $M = \pm\frac{\sqrt{\omega^4+\epsilon_0^4-2\omega^2\epsilon_0^2(2n+1)}}{2 \omega}$. 

Thus, finally, we end up with
\begin{align}
  \text{Re}\Big(\sigma_{xx}(\omega)\Big) & = 2e^2\frac{v_F}{\pi} \Big(f(- \frac{\epsilon_0^2+\omega^2}{2\omega}) - f(\frac{\epsilon_0^2+\omega^2}{2\omega}) + f( \frac{\epsilon_0^2-\omega^2}{2\omega}) - f( -\frac{\epsilon_0^2-\omega^2}{2\omega}) \Big) \sum_{n} \frac{k_{z,n+1}|-\omega^2+\epsilon_0^2(2n+1)| }{\omega\sqrt{\omega^4+\epsilon_0^4-2\omega^2\epsilon_0^2(2n+1)}} \nonumber \\
                         & = 2e^2\frac{v_F}{\pi}\Big(\frac{\sinh(\frac{\epsilon_0^2+\omega^2}{2\omega T})}{\cosh(\mu/T)+\cosh(\frac{\epsilon_0^2+\omega^2}{2\omega T})} - \frac{\sinh(\frac{\epsilon_0^2-\omega^2}{2\omega T})}{\cosh(\mu/T)+\cosh(\frac{\epsilon_0^2-\omega^2}{2\omega T})}\Big) \sum_{n} \frac{k_{z,n+1}|-\omega^2+\epsilon_0^2(2n+1)| }{\omega\sqrt{\omega^4+\epsilon_0^4-2\omega^2\epsilon_0^2(2n+1)}}.
\end{align}
\end{widetext}

This function has poles at $\omega=\epsilon_0(\sqrt{n+1}+\sqrt{n})$. Similarly, $\sigma_{yy}$ and $\sigma_{zz}$ can also be calculated.

\section{Location of poles}

The poles change their location as a magnetic field is introduced. This is shown in Fig. \ref{fig:poleshift} where the poles of only $\sigma_{yy}$ are compared. The continuous blue line depicts the case where $\textbf{B}=0$ and the dashed red line is for the case where $\ell_S/\ell_B=0.3$. For these two cases, the poles are expected at $2\sqrt{n}\epsilon_0$ and $2\sqrt{n}\epsilon_1$ respectively.
%\pagebreak

\end{document}